\documentclass[10pt,conference]{IEEEtran}

\usepackage[utf8]{inputenc}
\usepackage[T1]{fontenc}
\usepackage{soul}
\usepackage{xspace}
\usepackage{url}
\usepackage{graphicx}
\usepackage{multicol}
\usepackage[export]{adjustbox}
\usepackage{amsmath}
\usepackage{amssymb}
\usepackage{textcomp}
\usepackage{cleveref}
\usepackage{algorithm}
\usepackage[noend]{algpseudocode}
\usepackage{multirow}
\usepackage{booktabs}
\usepackage{subcaption}
\usepackage{listings}
\usepackage{breakurl}
\usepackage{titlesec}
\usepackage{xcolor}

\newcommand{\eg}{\emph{e.g.,}\xspace}

\newcommand{\psiminer}{\textsc{PSIMiner}\xspace}
\newcommand{\idea}{IntelliJ IDEA\xspace}
\newcommand{\platform}{IntelliJ Platform\xspace}
\newcommand{\astminer}{\emph{astminer}\xspace}

\newcommand{\secpart}[1]{\subsection{#1}}

\makeatletter
\newcommand{\linebreakand}{%
  \end{@IEEEauthorhalign}
  \hfill\mbox{}\par\mbox{}\hfill
  \begin{@IEEEauthorhalign}
}
\makeatother

\lstdefinelanguage{Kotlin}{
  keywords={fun, private},
  ndkeywords={extractFromVariable},
  sensitive=false,
  comment=[l]{//},
  morecomment=[s]{/*}{*/},
  morestring=[b]',
  morestring=[b]"
}

\titlespacing*{\section}{0pt}{2pt}{2pt}
\titlespacing*{\subsection}{0pt}{2pt}{2pt}

\title{\psiminer: A Tool for Mining \\ Rich Abstract Syntax Trees from Code}

\author{
    \IEEEauthorblockN{Egor Spirin}
    \IEEEauthorblockA{
        \textit{JetBrains Research}\\
        \textit{Higher School of Economics}\\
        Saint Petersburg, Russia \\
    }
    \and
    \IEEEauthorblockN{Egor Bogomolov}
    \IEEEauthorblockA{
        \textit{JetBrains Research}\\
        \textit{Higher School of Economics}\\
        Saint Petersburg, Russia \\
    }
    \and
    \IEEEauthorblockN{Vladimir Kovalenko}
    \IEEEauthorblockA{
        \textit{JetBrains Research}\\
        \textit{JetBrains N.V.} \\
        Amsterdam, The Netherlands \\
    }
    \and
    \IEEEauthorblockN{Timofey Bryksin}
    \IEEEauthorblockA{
        \textit{JetBrains Research}\\
        \textit{Saint Petersburg State University}\\
        Saint Petersburg, Russia \\
    }
    \linebreakand
    \IEEEauthorblockA{
        \{egor.spirin, egor.bogomolov, vladimir.kovalenko, timofey.bryksin\}@jetbrains.com
    }
}

\begin{document}
\maketitle

\begin{abstract}
The application of machine learning algorithms to source code has grown in the past years. Since these algorithms are quite sensitive to input data, it is not surprising that researchers experiment with input representations. Nowadays, a popular starting point to represent code is abstract syntax trees (ASTs).
Abstract syntax trees have been used for a long time in various software engineering domains, and in particular in IDEs. The API of modern IDEs allows to manipulate and traverse ASTs, resolve references between code elements, etc. Such algorithms can enrich ASTs with new data and therefore may be useful in ML-based code analysis.
In this work, we present \psiminer ---a tool for processing PSI trees from the \platform.
PSI trees contain code syntax trees as well as functions to work with them, and therefore can be used to enrich code representation using static analysis algorithms of modern IDEs.
To showcase this idea, we use our tool to infer types of identifiers in Java ASTs and extend the code2seq model for the method name prediction problem.
\end{abstract}

\section{Introduction}\label{sec:introduction}

Research applying machine learning (ML) to source code analysis and manipulation grows every year~\cite{Ferreira2019dl_in_se, Li2018dl_in_se, yang2020dl_for_se}. 
In the early days, most ML models treated code as plain text, trying out ideas from Natural Language Processing in the Software Engineering (SE) domain~\cite{Ernst2017nlp_for_se, Yalla2015nlp_and_se, Le2020dl_for_se}. 
A code fragment is indeed a text document, however, it also has a richer underlying structure brought by the syntax of the programming language and the semantics of the program. 
Several recently introduced approaches suggest that researchers can benefit from this additional information in the form of abstract syntax trees (ASTs) while building their models and tools, for example, by mining paths from ASTs~\cite{alon2018pathbased}, by augmenting ASTs with additional edges~\cite{Allamanis2017graphs}, by applying convolutions to trees~\cite{Mou2015tbcnn}, or by simply traversing trees to build a representation~\cite{Zhang2020retrieval}.
This helps to improve results in various SE tasks: generation of code comments~\cite{Shido2019ext_treelstm}, prediction of method names~\cite{Fernandes2018graphs}, type inference for dynamically typed languages~\cite{Allamanis2020typilus}, detection of code clones~\cite{Buch2019code_clones}, and others.

Moreover, the idea of using code structure and semantics (and syntax trees in particular) is not new: they have been studied and used in the developing compiler theory for half a century. 
More recently, this complex nature of code has been explored in the area of integrated development environments (IDEs).
Based on the syntax trees of code and their semantic analysis, modern IDEs provide such features as code completion, refactoring, searching code, exploring packages, running static analysis checks, and others~\cite{Murphy2006ide_usage}. 

Existing research involving ASTs mostly uses stand-alone parsers to build the ASTs (for instance, JavaParser~\cite{javaparser} for Java, TypedAST~\cite{typed_ast} for Python, Tree-sitter~\cite{tree-sitter} for C++, etc.), and then researchers are left alone to invent their own ways to enhance these ASTs and build derived representations of code.
With with work, we make a step towards leveraging long-established and constantly developing IDE capabilities of static analysis to extract rich representations of code. With this data, researchers can improve their ML pipelines without diving into complicated mechanisms of processing source code into a desirable format

To do that, we introduce \psiminer, a tool for processing internal representations from \platform~\cite{intellij_platform}---an open-source platform developed by JetBrains for building IDEs and other code-related tools. 
Program Structure Interface~\cite{psi}, or PSI, is an underlying infrastructure that enables parsing code files and creating syntactic and semantic code models. 
Most features built on top of \platform rely heavily on PSI trees which are the base of all the internal code structure. 
Apart from that, PSI contains interfaces to analyze and transform code, \eg resolve references between code elements, or refactor code. 

To show \psiminer in action, we demonstrate how it can be used to enrich ASTs by resolving types of identifiers in Java code. For variables in AST leaves, \psiminer tries to determine their Java types by calling the PSI API. To show the benefits of such an approach, we use the extracted data to extend code2seq~\cite{Alon2018code2seq}, a popular model for the method name prediction task.

\section{Background}\label{sec:background}
\newdimen\figrasterwd
\figrasterwd\textwidth

\begin{figure*}[ht]
    \centering
    \parbox{\figrasterwd}{
        \parbox{.4\figrasterwd}{
            \centering
            \subcaptionbox{An example code snippet.\label{fig:code_example}}{
                \includegraphics[width=0.5\hsize]{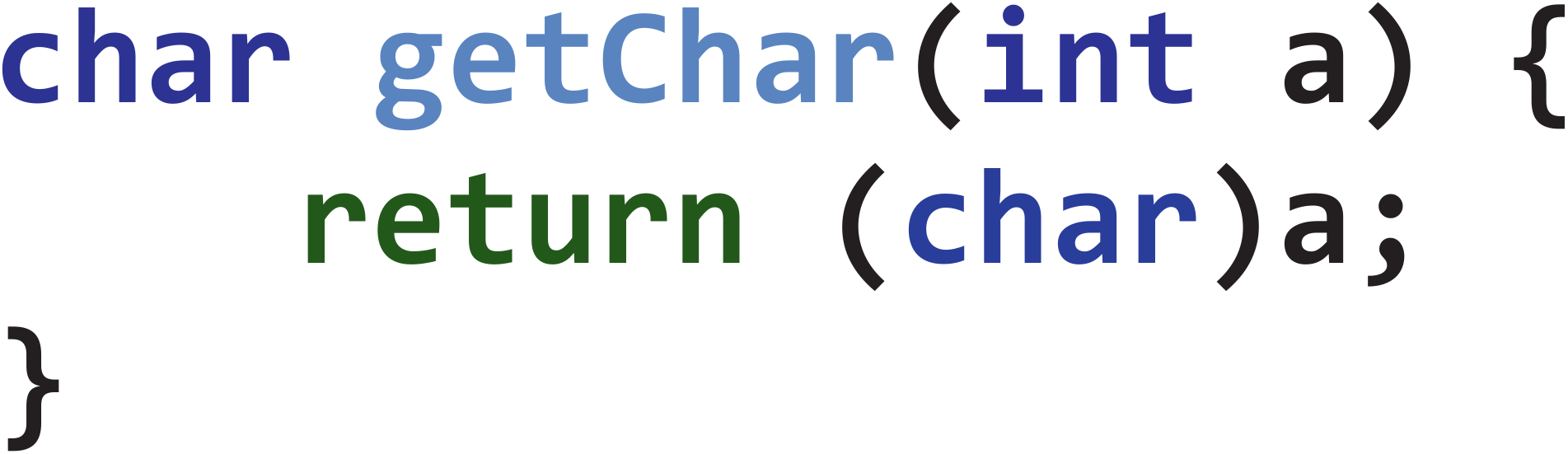}
            }
            \vskip3em
            \subcaptionbox{The snippet's abstract syntax tree. Intermediate nodes are colored red and leaf tokens, green.\label{fig:ast_example}}{
                \includegraphics[width=\hsize]{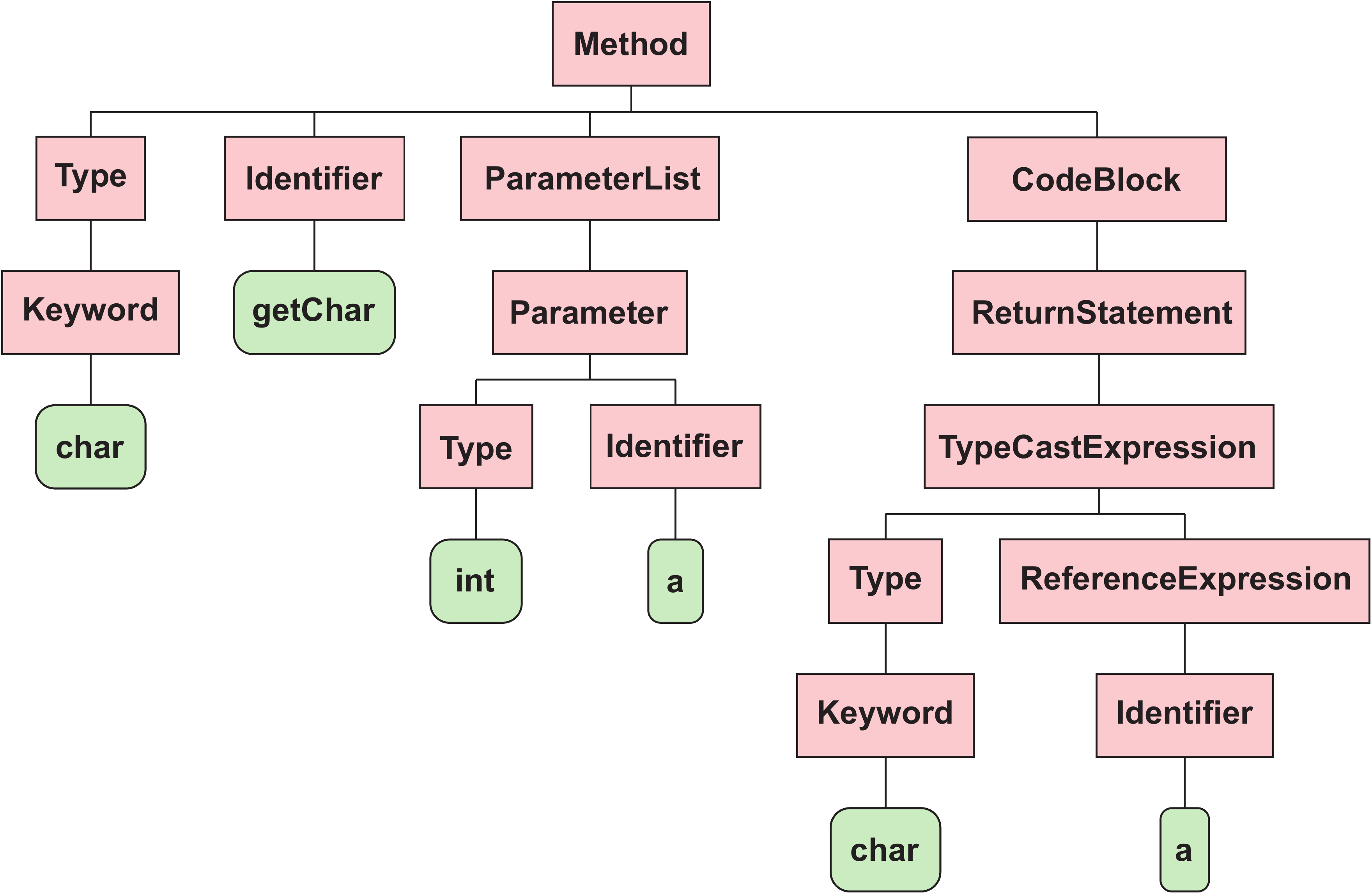}
            }  
        }
        \hskip1em
        \parbox{.575\figrasterwd}{%
            \subcaptionbox{The snippet's PSI tree without \texttt{PsiWhiteSpace} nodes. Red and green \\ nodes correspond to the AST. Blue nodes are created by PSI.\label{fig:psi_example}}{
                \includegraphics[width=\hsize]{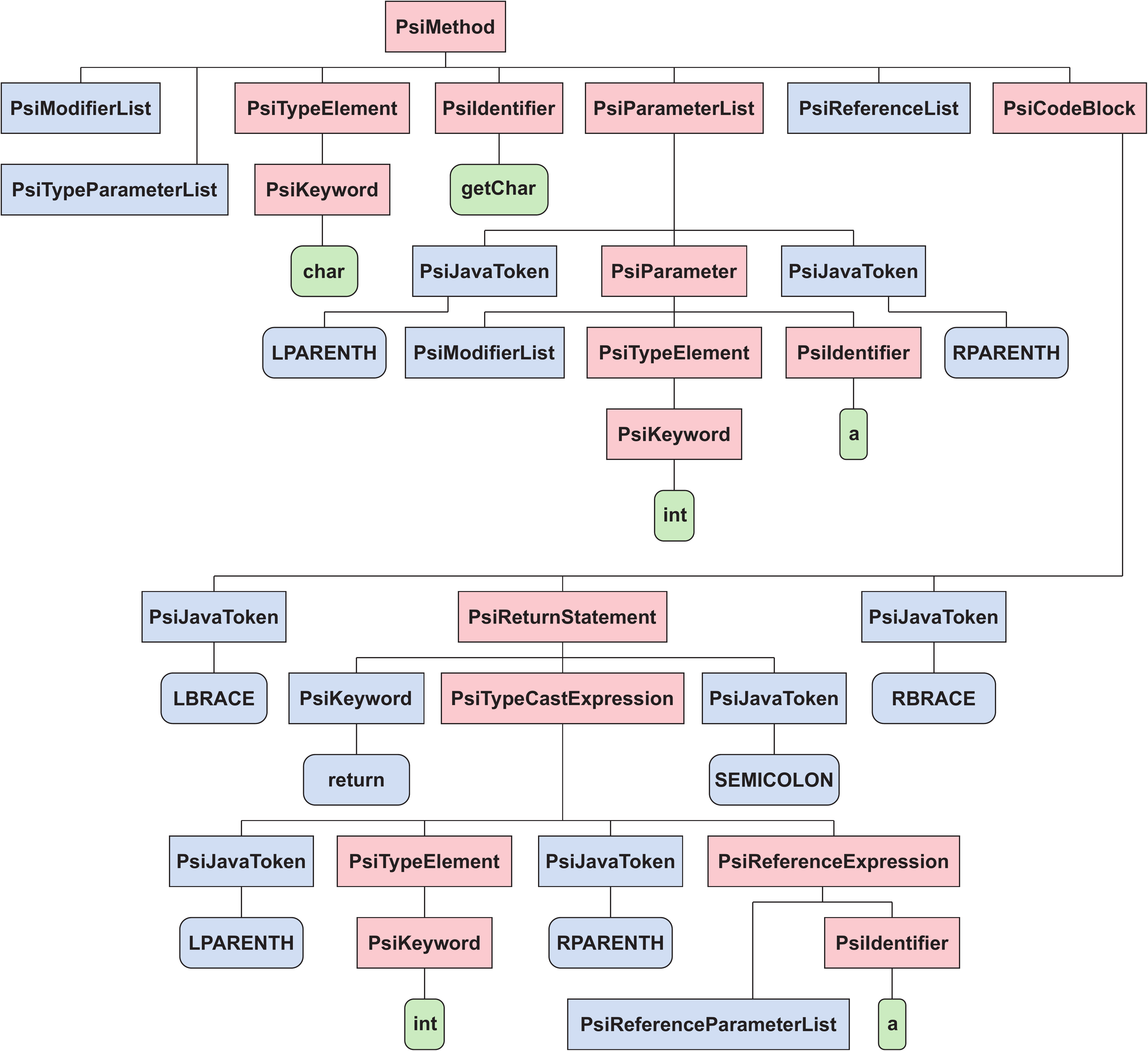}
            }

        }
    }
\caption{An example of a code snippet with corresponding trees}
\vspace{-0.5cm}
\end{figure*}

\secpart{Abstract Syntax Trees}

An abstract syntax tree (AST) is a tree representation of a program built during parsing.
Each node in an AST represents a code element: internal nodes represent operators (\eg conditions, loops, logical operations), and leaves stand for operands (\eg variables, literals, function calls).
An AST carries information about code tokens and their structural relations, which the parser builds based on the language grammar. 
In compilers, ASTs are usually enhanced with other types of information on later compilation stages (\eg type inference or other kinds of semantic analysis). 

\Cref{fig:code_example} presents a simple code fragment; an example of its AST is shown in \Cref{fig:ast_example}.

\secpart{Program Structure Interface}

One of the main features of \platform's internals is its infrastructure to analyze and manipulate code. 
Tools using it can rename identifiers, move methods between classes, search for specific tokens, etc. 
While doing this, it is critical to preserve the code semantics after each manipulation and perform such tasks as fast as possible. 
For this purpose, \platform introduces Program Structure Interface (PSI)---a combination of the syntax tree structure of code and the interfaces to interact with its parts. 
PSI trees act as concrete syntax trees: apart from the AST structure they also carry information about punctuation marks, braces, and other formatting.

Being a part of \platform, PSI trees are only accessible inside the IDE built on top of the platform, including plugins. 
A PSI tree contains all AST nodes, but also adds nodes introducing the internal structure, nodes for modifiers (even if they are not actually present), and other Java syntactic constructs.

\section{\psiminer internals}\label{sec:internals}

The purpose of \psiminer is to simplify working with the \platform's built-in code processing tools and, in particular, PSI trees. 
Since there is no obvious way to access PSI trees outside an IDE, we developed \psiminer as a plugin for \idea that extracts data from source code and stores it in a convenient format.
\psiminer runs \idea in the headless mode, without starting the IDE's GUI. This way our tool can be used from a command-line interface, making it easy to process datasets on remote servers.

\idea runs on JVM, so all of the API in \platform is in Java or Kotlin~\cite{kotlin}. We chose Kotlin to implement \psiminer.

\begin{figure*}[htbp]
    \centering
    \includegraphics[width=\textwidth]{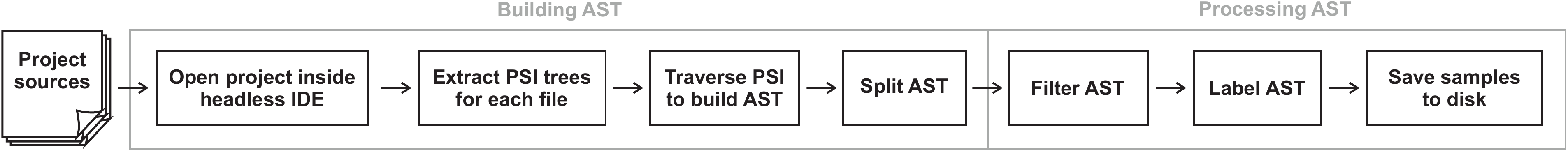}
    \caption{The pipeline of \psiminer. Step \textit{Traversing PSI to build AST} can be used to enrich ASTs with new information.}
    \label{fig:pipeline}
\vspace{-0.5cm}
\end{figure*}

\Cref{fig:pipeline} presents the pipeline of \psiminer. 
Its internals consist of two logical parts described in detail further in this section. 
The first part is responsible for parsing source code and building ASTs augmented with user-defined information. 
The second part of the tool handles processing of the extracted ASTs: filtering, labeling, and storing them on disk.

\secpart{Building ASTs}\label{sec:core}

The first step of the \psiminer's pipeline is reading input data. For our tool, it means loading code as a project inside the IDE. 
It is essential to import projects correctly, because PSI uses intra-project dependencies when, for example, resolving function declarations. 
SE datasets normally consist of large numbers of open-source projects collected, \eg from GitHub. As loading massive datasets all at once significantly increases memory consumption, \psiminer processes projects one by one, thus allowing PSI to resolve intra-project references with moderate memory usage. However, loading data at the project level is recommended but not required. \psiminer will work even with a single file as input.

The next step is filtering project files by their extension to keep only Java files. After that, \psiminer extracts PSI trees for the files by calling the \platform's API. 

The third step is traversing PSI trees to build code ASTs. This part of the tool takes previously extracted PSI trees as input and returns corresponding ASTs in accordance with the given parsing parameters. 
To produce an AST, \psiminer traverses PSI in the depth-first search order. 
For each node, it checks whether it should belong to the final AST by looking at its type.
\psiminer ignores some predefined types like \texttt{PsiWhiteSpace} because they do not carry any semantic information. 
Also it ignores nodes defined in its JSON configuration file, which can be useful, for example, to remove comments from code.
Further, the tool converts the nodes to an internal representation that contains the original PSI objects alongside with additional user-defined information obtained by calling the PSI API. 

In this work, we use an end-to-end example of inferring Java types for identifiers.
Their PSI nodes either already contain information about the identifier type, or it can be resolved by calling the appropriate API.
For example, for a variable, the type can be inferred with a simple function:
\begin{lstlisting}[language=Kotlin]
private fun extractFromVariable(
    node: PsiIdentifier
): String =
    node.parentOfType<PsiVariable>()
        ?.type
        ?.presentableText ?: NO_TYPE
\end{lstlisting}

The complete code for getting types for all identifiers fits into a small class.

The final step of the first part of \psiminer consists of splitting the ASTs according to the given level of granularity (file, class, or method).

\secpart{Processing ASTs}\label{sec:processing}

Having processed code using IDE capabilities, \psiminer stores the obtained data in a format appropriate for further application in an ML pipeline. In this second logical block of \psiminer, there are several steps for filtering, labeling, and storing data. All of them provide interfaces and, therefore, act as extension points for straightforward reusing to match an existing ML pipeline.

The task of extending \psiminer with new features requires only implementing an appropriate interface and does not need to change other parts of the pipeline. 
In the rest of this subsection, we describe such possible points of extension.

\textbf{Filter}. 
Implementing this interface allows to filter out ASTs based on user-defined conditions:
tree properties, {\eg} number of nodes or depth, or information from code, {\eg} checking for a particular modifier or even working with source code directly.

\psiminer already supports filtering out abstract and overridden methods, class constructors, as well as filtering by the AST size and the number of lines in the code fragment.

\textbf{Label extractor}.
This interface allows to declare what information from an AST should be treated as a label. For example, in the method name prediction problem, a method name is a label that the model should be trained to receive.
Firstly, the user should define the granularity level by assigning the enum variable (\eg class or method).
Secondly, the user should implement a function that accepts an AST and returns a pair: an extracted label and the corresponding AST. In our example, this function was used to extract method names and hide all mentions of them from the trees---ASTs may be changed during label extraction.

Currently, \psiminer can prepare datasets for method name prediction. 
Additionally, it can be quickly extended to solving other problems. For example, for a comment generation task, the label would be the text of the JavaDoc comment extracted from the tree.
On the other hand, if the task at hand requires only an AST without labels, this step may be skipped by using a dummy implementation of \textit{LabelExtractor}.

\textbf{Storage}.
This interface allows to choose an output format to match the requirements of the ML model. It defines several functions for processing data points into a required format with further saving on the disk and printing out dataset statistics in the end.

Currently, {\psiminer} supports storing datasets in two independent formats. First, as a JSONL file where each tree is saved in the same format as in the Python150k dataset~{\cite{raychev2016python150k}}.
As it is quite a popular dataset, we hope that providing output in the same format will help testing and further developing our approach.
Second, the output can be saved in a code2seq-compatible format which requires extracting a bag of paths from each tree and then storing them in a CSV-like file.
The output of the code2seq storage is compatible with the original implementation of code2seq.\footnote{\url{https://github.com/tech-srl/code2seq}}
To mine paths from each tree, we use {\astminer}~{\cite{kovalenko2019pathminer}}, a library designed to mine path-contexts and other derived data from ASTs.

\secpart{Supported Programming Languages}

In general, PSI is language-dependent. As \Cref{fig:psi_example} shows, a PSI tree contains language-specific nodes. 
For now, \psiminer supports only Java, one of the most popular programming languages. 
However, \platform supports a wide range of languages, providing different PSI variations for them. 
All PSI dialects share a large part of common code but differ in the nodes for syntax-specific tokens of a particular language. 
Thus, adding support for another language to \psiminer requires only to handle tree nodes specific to this language.
\section{\psiminer usage example}\label{sec:usage}

Since {\psiminer} is a tool for using static analysis algorithms from modern IDEs to prepare data for ML pipelines, we demonstrate an example of enhancing ASTs with identifier types to improve one of the existing method name prediction models called code2seq.

\secpart{code2seq}\label{sec:code2seq}

code2seq~\cite{Alon2018code2seq} is a neural network for generating sequences from snippets of code. 
For the name prediction task, these sequences are words that are concatenated in the end to form a method's name. 
The approach is based on path-based representation of code~\cite{alon2018pathbased}. 
As input, the model takes a set of path-contexts; each of them is a triple of two tokens from AST leaves and a path between them---the shortest sequence of internal AST nodes connecting two target leaf nodes. 

The encoder part builds separate embeddings for tokens and paths, and combines them into a single vector.
To build token embeddings, the model splits tokens into subtokens by CamelCase or snake\_case, converts subtokens into vectors via an embedding matrix $E^{token}$, and sums the vectors to get the token embedding.
For paths, code2seq first embeds the types of AST nodes into numerical vectors using another embedding matrix $E^{node}$ to get a sequence of vectors for each path.
After that, the sequence is passed through an LSTM, and the last state of this LSTM is treated as the path's embedding.
The last step of the encoder is aggregating the built vectors into a path-context embedding. 
For this, the model concatenates the vectors and passes them through a one-layer fully-connected neural network.

\secpart{Suggested Model Improvement}

To showcase \psiminer, we extend the code2seq model to handle type information for identifiers, which we can easily get from PSI trees when preparing the dataset.  

Within code2seq, we suggest handling types in the same way as the other parts of the path-context: embed the types into numeric vectors and concatenate them with vectors for tokens and paths. 
We embed types following the same algorithm as for identifier names: we split the type name into subwords, obtain embeddings for each of them separately, and then use the sum of these vectors to get an embedding for the whole type name.
After that we concatenate the type embedding with the embeddings of the tokens and the path before passing the result to a one-layer fully-connected neural network as described in \Cref{sec:code2seq}.

\secpart{Evaluation Results}

We reimplemented the code2seq model\footnote{\url{https://github.com/JetBrains-Research/code2seq}} and modified it as mentioned above.
We trained both models, original code2seq and its type-enhanced version, using the same path-contexts for fair comparison.
We evaluated both models on Java-small and Java-med datasets used in the original paper to evaluate code2seq for the method name prediction task. 
\Cref{table:results} presents the results. 
Although the original model achieved better precision on Java-med, the type-enhanced model achieved the best F1 score for both datasets.

\begin{table}[t]
\centering
\caption{Comparison of the original and the type-enhanced code2seq models on the method name prediction task.}
\begin{tabular}{lcccc}
\toprule
\textbf{Dataset}                     & \textbf{Model}    & \textbf{Precision} & \textbf{Recall}  & \textbf{F1 score} \\ \midrule
\multirow{2}{*}{Java-small} & Original & $38.76$   & $28.63$ & $32.93$   \\
                            & Type-enhanced    & $\textbf{41.68}$   & $\textbf{29.76}$ & $\textbf{34.73}$  \\
\midrule
\multirow{2}{*}{Java-med}   & Original & $\textbf{51.38}$ & $39.08$ & $44.39$ \\
                            & Type-enhanced    & $49.15$ & $\textbf{42.35}$ & $\textbf{45.5}$ \\ 
\bottomrule
\end{tabular}
\label{table:results}
\end{table}
\secpart{Data Leak}

The numbers in \Cref{table:results} are different from the numbers reported in the original paper by Alon et al.~\cite{Alon2018code2seq}.
We detected a data leak in the original dataset: to prevent the model from copying method names instead of predicting, the code2seq authors replaced the original names with special tokens in the method headers, but not in the method body where the names occur due to recursive calls. To address this issue, we traversed through the method body and replaced all recursive calls with another special token in the original datasets. We evaluated code2seq and our type-enhanced model on thus ‘cleaned’ datasets, code2seq receiving significantly lower scores.
When we do \textit{not} mask such recursive calls for Java-small, our implementation of the original model achieves the F1 score of roughly $41$, which is close to the score reported in~\cite{Alon2018code2seq}. 
The remaining minor difference in scores might be attributed to the difference in used parsers.

\section{Conclusion}\label{sec:conclusion}

In this work, we propose to enrich AST-based code representation with information available inside an IDE. 
To do that, we developed a tool called \psiminer.
While designing and implementing the tool, we kept two ideas in mind: straightforward access to \idea capabilities and convenient end-to-end creation of labeled datasets for use by ML models in the SE domain.

To showcase the usefulness of \psiminer, we enhanced ASTs with identifier types, which allowed us to improve the results of one of the existing models (code2seq) on the method name prediction problem.

Apart from ASTs, PSI trees contain much more information that can be used, \eg in building control flow or data flow graphs, mining contexts of variable or method usage, etc. These and other representations can be flexibly obtained using PSIMiner with minimal technical effort by working with extracted PSI or implementing interfaces presented in {\ref{sec:processing}}. {\psiminer} is only a prototype and we believe that the simple case show will motivate other researchers to experiment with more complex data that an IDE can provide and further improve code representation.
For that, we made all the source code publicly available.\footnote{\url{https://github.com/JetBrains-Research/psiminer}}

\bibliographystyle{IEEEtran}
\bibliography{IEEEabrv,paper}

\end{document}